\documentclass[10pt]{article}
\usepackage{amssymb}
\usepackage{amsfonts}
\usepackage{amsmath}
\usepackage{graphicx}
\usepackage{amsfonts}
\usepackage{amssymb}
\usepackage{epstopdf}
\usepackage{color}
\usepackage{float}

\usepackage[section]{placeins}

\usepackage{tikz}
\usetikzlibrary{arrows.meta, positioning}

\newcommand{\Z}{{\mathbb Z}}

\begin{document}
\thispagestyle{empty}

\begin{center}
\huge {A Stochastic ISCS Markov Model for Fake News Propagation
}

\vspace{.5cm}

\normalsize {\bf Carles
Rovira}

{\footnotesize \it Facultat de Matem\`atiques i Inform\`atica, Universitat de Barcelona, Gran
Via 585, 08007-Barcelona.

 {\it E-mail addresses}: 
Carles.Rovira@ub.edu}

\end{center}

\begin{abstract}
This paper studies the propagation of fake news through a stochastic rumor spreading model based on Markov chains. Inspired by classical epidemiological SIR models, we consider a generalization of the Daley-Kendall  framework for rumours that incorporates fact-checkers, following the Ignorant–Spreader–Checker-Stifler model introduced in \cite{Piqueira2020}. The model analyzes the influence of checkers on fake news dynamics. Numerical simulations are used to illustrate the behavior of the system and the impact of fact-checkers.

\end{abstract}



{\bf AMS 2000 MSC:} 92D30, 60J10, 60H10.

\renewcommand{\theequation}{1.\arabic{equation}}
\setcounter{equation}{0}
\section{Introduction}

Modern communications and technological advances have driven significant social progress. However, they have also facilitated the spread of misinformation.
Detecting fake news is a complex challenge. Microscopic approaches based on artificial intelligence can identify false or misleading content, but they often fail to capture the overall behavior of information diffusion. For this reason, since both human behavior and the spread of diseases arise from interactions among individuals, macroscopic approaches inspired by epidemiological models are frequently used to study how information spreads through social interactions.

One of the best-known epidemiological models is the SIR model  introduced by Kermack and McKendrick in \cite{KM}. The model classifies individuals into three compartments: Susceptible (S), Infected (I), and Recovered (R).

Inspired by the epidemiological SIR framework, Daley and Kendall proposed the Daley–Kendall (DK) rumor spreading model \cite{DK}, also known as the ISS model: Ignorant–Spreader–Stifler. In the ISS framework, the population is divided into three classes: ignorants, who have never heard the rumor; spreaders, who know and actively propagate it; and stiflers, who know the rumor but no longer spread it. The ISS model shares several similarities with the classical SIR model. In particular, the ignorant and spreader populations correspond respectively to the susceptible and infected classes of the SIR model. The key difference lies in the inclusion of the stifler class, which has no direct counterpart in the classical SIR formulation since stiflers inhibit further rumor propagation.

Several years later, Maki and Thompson introduced a modified version of the model, the Maki-Thompson (MT) model, in \cite{MT}, in which the behavior of stiflers differs slightly from that considered in the original DK formulation. In the DK model, a spreader becomes inactive after interacting with another spreader or a stifler, regardless of who initiated the contact. In the MT model, only the spreader who initiates such a contact becomes inactive, making the rumor typically persist longer than in the DK model.

The literature based on the Daley–Kendall (DK) and Maki–Thompson (MT) rumor-spreading models is extensive,  both deterministic and stochastic formulations, as well as continuous-time and discrete-time frameworks (see \cite{Belen} and the references therein).

In \cite{Piqueira2020}, where the ISS model is interpreted as a generalization of the classical DK model, the authors introduce a new category of individuals corresponding to fact-checkers, or checkers, leading to the Ignorant–Spreader–Checker–Stifler (ISCS) model. This additional compartment makes it possible to incorporate fact verification  and to study whether that may reduce the spread of fake news. Other relevant references that analyze the role of checkers include 
\cite{lotito2021realistic}, \cite{dicrescenzo2023stochastic}, and 
\cite{raponi2022fake}. On the other hand, several mathematical models have been proposed to study rumor propagation under different assumptions and social contexts, including skepticism in social networks, viral meme propagation, and limited information exchange (see \cite{WangMemes}  and \cite{Mengyun}).

The main objective of this work is to study the evolution of fake news by analyzing the influence of individuals who verify information, namely the checkers, following the approach introduced in \cite{Piqueira2020}. To this end, we consider a generalization of the Daley-Kendall model that incorporates the classes introduced in \cite{Piqueira2020}, adopting a stochastic framework based on Markov chains, following the epidemiological SIR model studied by Tuckwell and Williams. 

The originality lies in the development of a stochastic rumor-spreading framework that integrates fact-checking  within a Markovian structure. We provide a rigorous formulation that combines the Daley-Kendall model with an Ignorant–Spreader–Checker–Stifler structure in discrete time. The analysis considers  two complementary  regimes: multiple contacts per unit time and probabilistic single-contact dynamic.  We also note that, in addition to Markov chain techniques, we use approximations to deterministic models and to Galton–Watson branching processes.

The paper is organized as follows. Section 2 reviews a well-known epidemiological SIR model based on Markov chains. In Section 3, we introduce our ISCS model and we study its dynamics. Finally, Section 4 uses simulations to analyze the role of checkers in the spread of fake news.

\renewcommand{\theequation}{2.\arabic{equation}}
\setcounter{equation}{0}
\section{SIR type discrete time epidemic stochastic models}

Tuckwell and Williams in \cite{TW}
proposed a model  based on a discrete time Markovian approach. This model  is characterized by three main assumptions. First, the population is homogeneous, meaning that every individual has the same probability of becoming infected and the same probability of infecting others. Second, all infected individuals remain in class I for $r$ units of time. Finally, individuals either remain susceptible or follow the progression S→I→R.
They still assume that the population size is fixed and equal to $n$, and that the time
is discrete.
Furthermore, they assume:
\begin {enumerate}
\item \emph{Definition of a sick individual}:
given any individual $i$, with $i=1,.....n$, 
we define a stochastic process $Y^{i}=\{ Y^{i}(t), t=0,1,2,...\} $ such that 
$Y^{i}(t)=1$ if the individual is infectious at time $t$, otherwise $Y^{i}(t)=0$. 
The total number of infectious individuals at time $t\ge 0$ will be therefore equal to
$Y(t)= \sum_{i=1}^{n} Y^{i}(t)$.
\item \emph{Daily encounters}: each individual $i$, over $(t,t+1]$, 
will encounter a number of other individuals equal to 
$N$. 
\item \emph{Duration of the disease}: any individual remains infectious for $r$ consecutive days,
where $r$ is a positive integer. After this period, the individual recovers and becomes immune.
\item \emph{Contagious probability}: if an individual who has never been diseased up to and including time $t$,
encounters an individual in $(t,t+1]$ who is infectious at time $t$,
then independently of the results of other encounters, the encounter results in transmission of the disease 
with probability $p$.
\item \emph{Encounter probability}:
given $Y(t)=y$, the probability that a randomly chosen individual is infectious at time $t$ is given
by $y/n$.
\end{enumerate}
This model can be seen as a $(r+1)$-dimensional Markov chain with:
\begin{itemize}
\item $Y_l(t)$ the number of individuals who are infected at $t$ and have been infected for
exactly $l$ time units, with $l=0,1,2,...,r-1$;
\item $X(t)$ the number of susceptible individual at time $t$;
\item $Z(t)$ the number of individuals who were previously infected and are recovered at $t$.
\end {itemize}

\section{The ISCS model }

In this section, we introduce an extension of the SIR model proposed by Tuckwell and Williams. We consider two different situations. In the first one, each individual has N contacts during each unit of time. In the second one, each individual can have at most one contact, occurring with a given probability $p_c$. We present the first case in detail, which will allow us to derive the second case directly.

The model consists of four classes of individuals: ignorants (I), spreaders (S), checkers (C) and stiflers (S). Ignorants are individuals who have not heard the rumor or fake news. Checkers are individuals who have heard the rumor, verified that it is false, and communicate this verification whenever they interact with someone spreading it. Spreaders are individuals who believe the rumor and continue spreading it until they either meet  individuals who already know the rumor in $r$ different units of time or encounter a checker who convinces them that the information is false. Finally, stiflers are individuals who know the rumor but are no longer interested in spreading it and do not act as checkers.

\subsection{ Model description}

We assume that the population size is fixed and equal to $n$.  The time
is discrete. Furthermore, we can now define the model  in the following way:

\begin {enumerate}

\item \emph{Daily encounters}: each individual, over $(t,t+1]$, 
will have a number of encounters with other individuals equal to $N$. $N$ can either be greater than 1, or, if $N = 1$, the contact will occur with probability $p_c$.

\item \emph{Definition of a spreader individual}:
given any individual $i$, with $i=1,.....n$, 
we define a stochastic process $Y^{i}=\{ Y^{i}(t), t=0,1,2,...\} $ such that 
$Y^{i}(t)=1$ if the individual is a spreader at time $t$, otherwise $Y^{i}(t)=0$.
The total number of spreaders at time $t\ge 0$ will be therefore equal to
 $Y(t)=\sum_{i=1}^{n} Y^{i}(t)$.

\item \emph{Definition of a checker individual}:
given any individual $i$, with $i=1,.....n$, 
we define a stochastic process $V^{i}=\{ V^{i}(t), t=0,1,2,...\} $ such that 
$V^{i}(t)=1$ if the individual is a checker at time $t$, otherwise $V^{i}(t)=0$.
The total number of checkers at time $t\ge 0$ will be therefore equal to
 $V(t)=\sum_{i=1}^{n} V^{i}(t)$.

\item \emph{Entry to  stifler and checkers stages}: if an ignorant individual up to and including time $t$,
has an encounter with an individual in $(t,t+1]$ who is a spreader at time $t$,
the ignorant becomes a spreader with probability $p$ or a checker with probability $1-p$. If the ignorant has an encounter with  $j$ spreaders in $(t,t+1]$, then if at least one of these encounters turns them into a checker, they will become a checker.

\item \emph{Duration of the spreader stage}: An individual enters the spreader stage and remains in this state until they have an encounter   with a checker or he has encounters with non-ignorants in $r$ units of time.

\item \emph{Duration of the checker stage}: An individual enters the checker stage and remains in this state, unless they have an encounter with a  spreader in which case he transitions to the stifler state with probability $q$ or remains as a checker with probability $1-q$.

\item \emph{Encounter probability}: When encounters occur, all individuals have the same probability, whether they are ignorants, spreaders, checkers  or stiflers.

\end{enumerate}

We emphasize that, as in these types of models, the difference with the epidemiological model of Tuckwell and Williams lies in how individuals leave the spreader class (infected in the epidemiological model). In the case of rumor spreading, this occurs when they encounter a checker who already knows the rumor or when they have had contacts with individuals who already know it in $r$ units of time, whereas in the epidemiological case it is a matter of spending $r$ days in the infected state. On the other hand, we note that we always refer to contacts between individuals without considering who initiates the contact; therefore, we are working within the framework of the DK model.

\subsubsection{ Markovian model}

This ISCS model can be viewed as a Markovian model in two different ways: either by considering the entire population globally or by studying the dynamics of each individual.

\medskip

\noindent {\bf First Markovian model:}  Let
\begin{itemize}
\item $X(t)$ the number of ignorant individuals  at time $t$;
\item $Y_{l}(t)$ the number of spreader individuals who have been in the spreader state during the time in which they have been in contact  non-ignorant individuals who are not checkers in $l$ units of time, with $l=0,1,2,\ldots,r-1$;
\item $V(t)$ the number of individuals who  are in the checker stage at $t$;
\item $Z(t)$ the number of individuals who  are in the stifler stage at $t$.
\end {itemize}
Notice that  $Y(t)= \sum_{j=0}^{r-1} Y_{j} (t)$ the total number of spreader  individuals at time $t$ . 

This model can be seen as a $(r+2)$-dimensional Markov chain, that is,
\begin{equation*}
V(t)=(X(t),Y_{0}(t),Y_{1}(t),
\ldots , Y_{r-1}(t),V(t)),  t=0,1,2,... 
\end{equation*}
forms a Markov chain
with state space 
\begin{eqnarray*}
S(n,r)=
\big\{(x,y_{0},\ldots,y_{r-1},v): x,v,y_{i} \in \Z_+, \ \\  \mbox{for} \ i=0, \ldots, r-1, \mbox{and} \ 
x+v+\sum_{i=0}^{r-1} y_{i} \le n \big\} .
\end{eqnarray*}

\medskip

\noindent {\bf  Second Markovian model:}
In addition to the process $Y^{i}=\{ Y^{i}(t), t=0,1,2,...\} $, 
we can define in a similar manner the process $X^{i}=\{ X^{i}(t), t=0,1,2,...\} $, for $i=1,\ldots,n$, 
which indicates whether individual $i$ is an ignorant or not, and the process
\[
Z^i(t)=1-X^{i}(t)-Y^{i}(t)-V^{i}(t)
\]
which indicates if the individual $i$ is in the stifler stage. 
We immediately get
$$
X(t)=\sum _{i=1}^{n}X^i(t), \, Y(t)=\sum _{i=1}^{n}Y^i(t) ,\,V(t)=\sum _{i=1}^{n}V^i(t) ,\,Z(t)=\sum _{i=1}^{n}Z^i(t) .
$$
We can also consider the processes $Y_{0}^i,Y_{1}^i, ..., Y_{r-1}^i$, where $i=1,2,...,n$,
and $Y_{k}^i(t)=1$ if the individual $i$ at time $t$ is in spreader stage and he has had contacts with spreaders or stiflers in $k$ units of time, zero otherwise. 
Then
$$
Y^{i}(t)= \sum _{k=0}^{r-1}Y_{k}^i(t).
$$

With these definitions we can consider a new Markovian model
$$
\textbf{M}(t)=[X^i(t), Y_{0}^i(t), Y_{1}^i(t),. . ., Y_{r-1}^i(t), V^i(t),  i=1,2, . . ., n] 
$$
whose state space is now
\[
\begin{array}{rl}
S_1(n,r) = & \big\{ 
(x^1,\ldots,y^1_{r-1},v^1,\ldots, x^{n},\ldots,y^{n}_{r-1},v^n )  \\
& \ \ \       \in \{0,1\}^{n(r+2)}: 
   x^i+v^i+\sum_{k=0}^{r-1} y_{k}^i \le 1 \
\mbox{for} \ i=1, \ldots, n
\big\} .
\end{array}
\]

\subsection{ Case $N>1$}

Furthermore, we can now study the model with $N>1$ contacts in the following way:
\subsubsection{The evolution of an individual }\label{probtran}

In this subsection, we will study the transition probabilities of an individual. Let us now fix the individual $i$ and study the process
\[
M_{i}(t)=[X^i(t), Y_{0}^i(t), Y_{1}^i(t),. . ., Y_{r-1}^i(t),V^i(t)],
\]
where $X^i(t)+ Y_{0}^i(t)+ Y_{1}^i(t)+ \cdots+Y_{r-1}^i(t)+V^i(t) \le 1$ and set
$$Z^i(t)=1-(X^i(t)+ Y_{0}^i(t)+ Y_{1}^i(t)+ \cdots+Y_{r-1}^i(t)+V^i(t)).$$ Clearly, $Z^i(t)=1$ if and only if individual $i$ belongs to the stifler class at time $t$.

The first  interesting case occurs $X^i(t)=1$
and we have to calculate the probability that this
ignorant  individual becomes a checker or a spreader for the first time at $t+1$.
Assuming $n$ much greater than $N$, 
we can approximate the probability of meeting exactly $j$ spreader individuals
with the binomial law, obtaining
\begin{equation}
\label{appro}
P_{j}^{i}(y) \approx \binom{N}{j} \Bigl(\frac{y}{n-1}\Bigr)^j \Bigl(1-\frac{y}{n-1}\Bigr)^{N-j},
\end{equation}
for $j\in\{0,\ldots , N\}$ and zero otherwise,
where $y=Y(t)$ is the total number of spreader individuals at time $t$. 

Observe that an ignorant individual who has contacts with $j$ spreaders will become a spreader with probability $p^j$ or a checker with probability $1-p^j$. Thus, the probability that the ignorant becomes a spreader will be
\begin{equation}\nonumber
\begin{split}
p_{i.sp}(y)&:=
P(Y_{0}^i(t+1)= 1|X^i(t)=1,Y(t)=y)=\sum_{j=1}^{N} p^j P_{j}^i(y) \\
                                                                                 &\approx \Bigl(1-\frac{(1-p)y}{n-1}\Bigr)^{N} - \Bigl(1-\frac{y}{n-1}\Bigr)^{N}
\end{split}                                                          	
\end{equation}
using the approximation (\ref{appro}). Moreover, following the same ideas,  the probability that the ignorant becomes a  checker will be
\begin{equation}\nonumber
\begin{split}
p_{i.c}(y)&:=
P(V^i(t+1)= 1|X^i(t)=1,Y(t)=y) =\sum_{j=1}^{N} (1-p^j) P_{j}^i(y) \\
                                                                                 &\approx1-   \Bigl(1-\frac{(1-p)y}{n-1}\Bigr)^{N}.
\end{split}                                                          	
\end{equation}

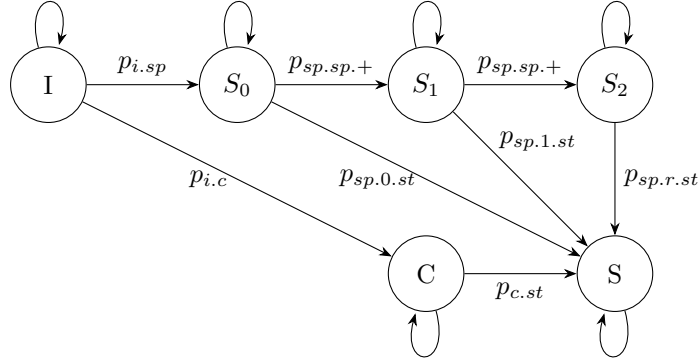
\begin{figure}[h]
\centering

\begin{tikzpicture}[
    node distance=2.5cm,
    state/.style={draw, circle, minimum size=1cm},
    >=Stealth
]

\node[state] (I) {I};
\node[state] (S0) [right of=I] {$S_0$};
\node[state] (S1) [right of=S0] {$S_1$};
\node[state] (S2) [right of=S1] {$S_2$};
\node[state] (C) [below of=S1] {C};
\node[state] (R) [below of=S2] {S};

\draw[->] (I) -- node[midway, above] {$p_{i.sp}$} (S0);
\draw[->] (S0) -- node[midway, above] {$p_{sp.sp.+}$} (S1);
\draw[->] (S1) -- node[midway, above] {$p_{sp.sp.+}$} (S2);

\draw[->] (I) -- node[midway, left] {$p_{i.c}$} (C);
\draw[->] (C) -- node[midway, below] {$p_{c.st}$} (R);

\draw[->] (S0) -- node[midway, left] {$p_{sp.0.st}$} (R);
\draw[->] (S1) -- node[midway, right, yshift=5mm, xshift=-4mm] {$p_{sp.1.st}$} (R);
\draw[->] (S2) -- node[midway, right] {$p_{sp.r.st}$} (R);

\draw[->] (I) edge[loop above] node {$$} (I);
\draw[->] (S0) edge[loop above] node {$$} (S0);
\draw[->] (S1) edge[loop above] node {$$} (S1);
\draw[->] (S2) edge[loop above] node {$$} (S2);
\draw[->] (C) edge[loop below] node {} (C);
\draw[->] (R) edge[loop below] node {$$} (R);

\end{tikzpicture}

{\caption{State transition diagram of the ISCS model with $r=3$ for an individual. $S_k$ denotes the class of spreaders who have  had contact with  spreaders or stiflers $k$ units of time.}}
\label{fig:iscr}
\end{figure}

Another  interesting case is when $Y_{k}^i(t)=1$, with $k \in \{0,\ldots,r-2\}$,
and we have to calculate the probability that this
spreader individual becomes a stifler at $t+1$, that is $Y_{k+1}^i(t+1)=0$ and $Y_{k}^i(t+1)=0$.
Clearly, since this only happens when they come into contact with a checker, using the same argument,
\begin{equation}\nonumber
\begin{split}
p_{sp.k.st}(v)&:=P(Z^i(t+1)= 1 |Y_{k}^i(t)=1,  V(t)=v )   \\
                                                                                 &\approx1- \Bigl(1-\frac{v}{n-1}\Bigr)^{N},
\end{split}                                                          	
\end{equation}
for $k \in \{0,\ldots,r-2\}$. 

The individual may also remain a stifler while having spread the rumor one more time; that is, among the $N$ contacts, at least one was with a non-ignorant individual who was not a checker.
\begin{equation}\nonumber
\begin{split}
p_{sp.sp.+}(x,v)&:=P(Y_{k+1}^i(t+1)= 1 |Y_{k}^i(t)=1, X(t)=x, V(t)=v )   \\
                                                                                 &\approx \sum_{j=1}^N  \binom{N}{j} \Bigl(\frac{n-x-v-1}{n-1}\Bigr)^j \Bigl(\frac{x}{n-1}\Bigr)^{N-j}, \\
                                                                                 &= \Bigl(1-\frac{v}{n-1}\Bigr)^{N}   - \Bigl(\frac{x}{n-1}\Bigr)^{N}          ,
\end{split}                                                          	
\end{equation}
for $k \in \{0,\ldots,r-2\}$, where we have used that the probability of contacting $j$ non-ignorant non-checkers and $N-j$ ignorants can be approximated by
$$\binom{N}{j} \Bigl(\frac{n-x-v-1}{n-1}\Bigr)^j \Bigl(\frac{x}{n-1}\Bigr)^{N-j}.$$

We must also compute the probability that a spreader who has already spread the rumor $r-1$ times moves to the stifler class. This will occur if they have contact with at least one non-ignorant individual. Thus:
\begin{equation}\nonumber
\begin{split}
p_{sp.r.st}(x)&:=P(Z^i(t+1)= 1|Y_{r-1}^i(t)=1, X(t)=x )   \\
                                                                                 &\approx1- \Bigl(\frac{x}{n-1}\Bigr)^{N} .  
\end{split}                                                          	
\end{equation}

Finally, we have to calculate the probability that a
checker individual at time $t$ becomes a stifler at $t+1$, that is $V_{k+1}^i(t+1)=0$ and $Z^i(t+1)=1$.
Using again the same arguments we can see
$$
p_{c.st}(y):=P(Z^i(t+1)= 1|V^i(t)=1, Y(t)=y)   \approx 1- \Bigl(1-q\frac{y}{n-1}\Bigr)^{N} .
$$

\subsubsection{ Following the evolution of the fake news}\label{evol}

Let us come back to the initial Markov chain. Assume, that for some $t$ we know
\begin{equation*}
W(t)=(X(t),Y_{0}(t),Y_{1}(t),
\ldots , Y_{r-1}(t),V(t)).
\end{equation*}
Let us recall that   $Y(t)= \sum_{j=0}^{r-1} Y_{j} (t)$ is the total number of spreader individuals at time $t$.
We need to  introduce new notation: for $k \in \{0,\ldots,r-2\}$, we define:
\begin{itemize}
\item $Y_k^{{new}}(t)$ as the number of spreaders individuals who, up to time $t-1$, had contacts with non-ignorant non-checkers over $k-1$ time units and have had  new contacts with a non-ignorant non-checker in $(t-1,t]$;
\item $Y_k^{{cont}}(t)$ as the number of spreaders individuals who, up to time $t-1$, had contacts with non-ignorant non-checkers over $k$ time units and have not had any new contact with a non-ignorant non-checker in $(t-1,t]$.
\end{itemize}
Similarly, we can define $V^{{new}}(t)$ and $V^{{cont}}(t)$. We also define $Z^{{new},k}(t)$, where the index $k$ indicates the new individuals entering the stifler class at time $t$ who were stiflers at time $t-1$ and are counted in $Y_k(t-1)$.

We can now compute, using the transition probabilities computed in the previous subsection, what happens to the population at time $t+1$ if we know its state at time $t$. 

First, we have the conditional law of the population that is still ignorant  and the new spreader $Y_{0}^{new}(t+1)$ and new checkers individuals $V^{new}(t+1)$ at time $t+1$
\begin{equation}\nonumber
\begin{split}
&(X(t+1),Y_{0}^{new}(t+1),V^{new}(t+1))_{ |X(t)=x,  Y(t)=y}  \\
                                                                                 & \qquad \sim  \text{Mul} (x;1-p_{i.sp}(y)-p_{i.c}(y),p_{i.sp}(y),p_{i.c}(y)).
\end{split}                                                          	
\end{equation}

Let us  now look at what happens to individuals who are in the spreader stage and have had  contacts in $k$ units of time, with $k \in \{0,\ldots, r-2\}$, with non-ignorant individuals who are not checkers. They may either move directly to the stifler group, remain as spreaders with the same number of contacts with non-ignorant individuals who are not checkers, or continue as spreaders but now with contacts in $k+1$ units of time. Thus, for $k \in \{0,\ldots,r-2\}$
\begin{equation}\nonumber
\begin{split}
&(Y_{k+1}^{new}(t+1),Y_{k}^{cont}(t+1),Z^{new,k}(t+1))_{ |X(t)=x, Y_{k}(t)=y_{k},   V(t)=v}  \\
& \qquad \sim  \text{Mul} (y_k;p_{sp.sp.+}(x,v),1-p_{sp.sp.+}(x,v)-p_{sp.k.st}(v),p_{sp.k.st}(v)).
\end{split}                                                          	
\end{equation}
In contrast, individuals who have already had $r-1$ contacts may either remain with the same number of contacts or move to the stifler class with probability $p_{sp.r.st}$, so that
$$  Y_{r-1}^{cont}(t+1)_{|Y_{r-1}(t)=y_{r-1}, X(t)=x} \sim \text{Bin} (y_{r-1},1-p_{sp.r.st}(x)).$$
Evidently, for $k \in \{0,\ldots,r-1\}$,
$$Y_{k}(t+1)=Y_{k}^{cont}(t+1) + Y_{k}^{new} (t+1).$$

Moreover, if $Y(t)=y$, the number of checkers that remain in the checker class follows
$$  V^{cont}(t+1)_{|V(t)=v,  Y(t)=y} \sim \text{Bin} (v,1-p_{c.st}(y))$$
and clearly
$$V(t+1)=V^{cont}(t+1)+V^{new}(t+1).$$

Finally, the number of individuals in the stifler class at time $t+1$ will be
$$Z(t+1) =n - X(t+1)-Y(t+1)-V(t+1),$$
or equivalently
$$Z(t+1)=Z(t)+\sum_{k=0}^{r-2}Z^{new,k}(t+1)+(V(t)-V^{cont}(t+1))+(Y_{r-1}(t)-Y_{r-1}^{cont}(t+1)).$$

 \subsubsection{Early-stage propagation of the fake news}
In this subsection we will study the propagation starting from a population of ignorants. We will consider, for $n$ large, the densities
\[
 \tilde{x}(t)=\frac{X([nt])}{n},\qquad
\tilde{y}(t)=\frac{Y([nt])}{n},\qquad
\tilde{v}(t)=\frac{V([nt])}{n}.
\]

As we have seen, the probability that an ignorant individual becomes a spreader at time $t$ if $\tilde{y}(t)=\tilde{y}$ is
\[
(1-(1-p)\tilde{y})^{N} - (1-\tilde{y})^{N}.
\]
For $\tilde{y}$ small, using the first-order Taylor expansion, $
(1-\tilde{y})^N = 1 - N\tilde{y} + \mathcal{O}(\tilde{y}^2)$, 
we get that the
probability of becoming a spreader can be approximated by $ 1-N(1-p)\tilde{y} -(1-N\tilde{y}) = Np \tilde{y}.$
Therefore, the total rate of creation of new spreaders is
\[
\mathcal{F}(\tilde{y}) \approx  \tilde{x} (p N \tilde{y}),
\]
and near the initial equilibrium $\tilde{x} \approx 1$, we obtain that
\[
\mathcal{F}(\tilde{y}) \approx p N \tilde{y}.
\]
Since at early stages ($\tilde{y} \approx 0$, $\tilde{v} \approx 0$), interactions with non-ignorants
are negligible, there will be no removal of spreaders.
Thus, if we consider the linearized dynamics we would obtain
\[
\dot{\tilde{y}} =
  p N \tilde{y}.
\]
Then, the condition for initial growth of the rumor is $pN>1$. 
This quantity can be considered as the 
Basic reproduction number $R_0$ in this framework.

Another way to study the early-stage behavior is by using branching processes.
In this early stage, the number of spreaders is very small compared with the population size; that is,
$
X(t)\approx n, \quad V(t)\approx 0,
$
and interactions among spreaders or between spreaders and checkers are negligible.
Under these assumptions, each spreader evolves approximately independently, and the propagation process can be described by a Galton--Watson branching process with offspring distribution
$
\xi \sim \mathrm{Bin}(N,p).$
Indeed, each spreader performs $N$ contacts during one unit of time. For each contact
 the contacted ignorant becomes a spreader with probability  $p$ or
 the contacted ignorant becomes a checker with probability $1-p$. 

Thus, the offspring generating function is
$
G(s)=\mathbb{E}[s^\xi]=(1-p+ps)^N.
$ and  the mean number of offspring is
$m=\mathbb{E}[\xi]=Np,
$
which is the quantity we have denoted by \(R_0\).
From the classical theory of Galton--Watson branching processes, it follows that:
\begin{itemize}
\item if \(R_0 \le 1\), extinction occurs with probability one (the fake news dies out);
\item if \(R_0 = Np > 1\), the extinction probability $s$ is the smallest solution in \([0,1]\) of
\[
s=(1-p+ps)^N.
\]
Consequently, the probability of a major rumor outbreak (i.e., that the fake news does not die out) is
$
1 - s.
$
\end{itemize}

Thus, increasing the number of contacts or the spreading probability makes
the rumor more likely to propagate and we have a phase transition. The condition $pN > 1$ defines
a critical regime separating extinction and large-scale fake news propagation. We observe that $p$ appears, the probability that controls whether a non-ignorant becomes a stifler or a checker. Evidently, if one wants to control fake news, the more contacts there are, the more checkers are needed.

\subsubsection{Early-Stage propagation  with fact-checkers }

Now, we would like to refine the early-stage analysis of the ISCS model by incorporating the effect of fact-checkers. 
We consider now that:
rhe number of spreaders is small compared to the population size $n$,
the number of checkers is small but nonzero,
the population is mostly composed of ignorants and
interactions among spreaders are negligible.

Under these assumptions, following the ideas of the previous subsection, each spreader evolves approximately independently and  fake-news propagation process can be approximated by a Galton--Watson branching process with offspring distribution
$
\xi $ where $\xi_{|A^c} \sim \mathrm{Bin}(N,p)$ with $$A:=\{ \text{a spreader is neutralized before producing offspring}\}.$$
Let $\tilde v$ denote the proportion of checkers in the population at an early stage. 
Thus, the probability of encountering at least one checker is
$
1 - (1 - \tilde v)^N.
$
and we can obtain the approximation
\[
P(A) \approx \left(1 - (1 - \tilde v)^N\right)\approx  N \tilde v.
\]
where we have used, 
for small $\tilde v$, the first-order Taylor expansion,
$
(1 - \tilde v)^N \approx 1 - N \tilde v
$.
In the stage, the rates of creation of spreaders and checkers are proportional to
$
N\tilde p y $ and $N(1-\tilde p) y,$
respectively, where $\tilde y$ denotes the proportion of spreaders,
which implies
\[
\tilde v \approx \tilde y \frac{1-p}{p}.
\]
Since $\tilde  y$ is small but positive, we obtain
\[
P(A) \approx  N \tilde y \frac{1-p}{p}.
\]

Thus, we can consider as a  reproduction number 
\[
R_{y}  = \mathbb{E}[\xi] \approx Np \left(1 -  N\tilde  y \frac{1-p}{p} \right).
\]
Then the fake news propagation process can be approximated by a branching process with reproduction number $R_y$.
Thus
if $R_{y} \leq 1$, the fake news becomes extinct with probability one and
if $R_{y} > 1$, the fake new survives with positive probability.

So we have that
the effect of fact-checkers is nonlinear, through the factor $\frac{1-p}{p}$,
reducing $p$  decreases $R_y$ and
the impact of fact-checkers is amplified by the number of contacts $N$.

\subsection{ Case $N=1$}

We now consider  the case in which the effective number of contacts is less than 1, that is, we assume that each individual has at most one contact per unit time, occurring with probability $p_c.$ 

\subsubsection{ Following the evolution of the fake news}

Since the model has similarities with the case $N>1$, we begin by adapting the transition probabilities computed in Subsection \ref{probtran} for the case $N=1$. As the method is the same, we only present the results. We then apply these new transition probabilities to study the evolution of fake news in a manner analogous to that carried out in Subsection \ref{evol} for the case $N>1$. 

We therefore begin with the transition probabilities. Notice that all transition probabilities must now be multiplied by the contact probability $p_c.$  Since \(N=1\), we obtain
\begin{equation}\nonumber
\bar p_{i.sp}(y):\approx p_c\Bigl( \Bigl(1-\frac{(1-p)y}{n-1}\Bigr) - \Bigl(1-\frac{y}{n-1}\Bigr)\Bigr)
= p_c p \frac{y}{n-1}.
\end{equation}
Similarly,
\begin{equation}\nonumber
\bar p_{i.c}(y):= 
p_c \Bigl(1- \Bigl(1-\frac{(1-p)y}{n-1}\Bigr)\Bigr)
= p_c (1-p)\frac{y}{n-1}.
\end{equation}
For \(k \in \{0,\ldots,r-2\}\), we have
\begin{equation}\nonumber
\bar p_{sp.k.st}(v) \approx p_c \frac{v}{n-1}.
\end{equation}
In the same way,
\begin{equation}\nonumber
\bar p_{sp.sp.+}(x,v) \approx p_c \Bigl(1-\frac{v+x}{n-1}\Bigr),
\qquad k \in \{0,\ldots,r-2\}.
\end{equation}
We also compute the probability that a spreader who has already had \(r-1\) contacts with non-ignorant, non-checker individuals transitions to the stifler class. This occurs if the individual meets a non-ignorant individual, hence
\begin{equation}\nonumber
\bar p_{sp.r.st}(x)\approx p_c \Bigl(1- \Bigl(\frac{x}{n-1}\Bigr)\Bigr).
\end{equation}
Finally, we compute the probability that a checker individual at time \(t\) becomes a stifler at time \(t+1\). Using the same arguments, we obtain
\[
\bar p_{c.st}(y) \approx p_c \Bigl(1- \Bigl(1-q\frac{y}{n-1}\Bigr)\Bigr)
= p_c q \frac{y}{n-1}.
\]
\subsubsection{Approximation to a Deterministic Model}

For this case, we can approximate this stochastic model by a deterministic one. Our goal is to show that we obtain a model similar to the well-known deterministic models for the study of rumors.

Let $\mathcal{F}_t$ denote the expectation conditioned on the natural filtration. Observe first that we have
\[
\begin{aligned}
&\mathbb{E}[X(t+1)-X(t)\mid\mathcal F_t] = -X(t)\bigl(\bar{p}_{i,sp}(Y(t))+\bar{p}_{i,c}(Y(t))\bigr)\\
& \qquad= -X(t)\,p_c\,\frac{Y(t)}{n-1},\\[6pt]
&\mathbb{E}[Y_0(t+1)-Y_0(t)\mid\mathcal F_t]\\
&\qquad=
X(t)\bar{p}_{i,sp}(Y(t))
-
Y_0(t)\bigl(\bar{p}_{sp.sp.+}(X(t),V(t))+ \bar{p}_{sp.0.st}(V(t))\bigr)\\
&\qquad=
p_c p X(t)\frac{Y(t)}{n-1}
-
p_c Y_0(t)\Bigl(1-\frac{X(t)}{n-1}\Bigr),\\
& {\rm for}  \,k \in \{0,\ldots,r-2\},\\
&\mathbb{E}[Y_k(t+1)-Y_k(t)\mid\mathcal F_t]\\
&\qquad=
Y_{k-1}(t)\bar{p}_{sp,sp,+}(X(t),V(t))
-
Y_k(t)\bigl(\bar{p}_{sp.sp.+}(X(t),V(t))-\bar{p}_{sp.k.st}(V(t))\bigr)\\
&\qquad=
p_c Y_{k-1}(t)\Bigl(1-\frac{V(t)+X(t)}{n-1}\Bigr)
-
p_c Y_k(t)\Bigl(1-\frac{X(t)}{n-1}\Bigr), \\
&\mathbb{E}[Y_{r-1}(t+1)-Y_{r-1}(t)\mid\mathcal F_t]\\
&\qquad=
Y_{r-2}(t)\bar{p}_{sp,sp,+}(X(t),V(t))
-
Y_{r-1}(t)\bar{p}_{sp,r,st}(X(t))\\
&\qquad=
p_c Y_{r-2}(t)\Bigl(1-\frac{V(t)+X(t)}{n-1}\Bigr)
-
p_c Y_{r-1}(t)\Bigl(1-\frac{X(t)}{n-1}\Bigr),\\
&\mathbb{E}[V(t+1)-V(t)\mid\mathcal F_t]
=
X(t)\bar{p}_{i,c}(Y(t))
-
V(t)\bar{p}_{c,sp}(Y(t))\\
&\qquad =
p_c(1-p)X(t)\frac{Y(t)}{n-1}
-
p_c q V(t)\frac{Y(t)}{n-1}.
\end{aligned}
\]

Using again  the densities as in the case $N>1$
\[
\tilde{x}(t)=\frac{X([nt])}{n},\qquad
\tilde{y}_k(t)=\frac{Y_k([nt])}{n},\qquad
\tilde{v}(t)=\frac{V([nt])}{n}.
\]
and dividing the balance equations by $n$ and letting $n\to\infty$, we can identify the expected discrete increment with the time derivative:
\[
\mathbb{E}\Big[\frac{X([n(t+1)])-X([nt])}{n}\Big]   =  \mathbb{E}[\tilde{x}(t+1)-\tilde{x}(t)] \;\longrightarrow\; \dot{\tilde{x}}(t),
\]
and analogously for all variables. We thus obtain the following system of ODEs:
\begin{eqnarray*}
\dot{\tilde{x}}&=&
- p_c  \tilde{x} \tilde{y}
\\
\dot{\tilde{y}}_0 &=
&
p_c p \tilde{x} \tilde{y}
-
p_c \tilde{y}_0 (1-\tilde{x})
\\
\dot{\tilde{y}}_k &=
&
p_c \tilde{y}_{k-1}(1-\tilde{x}-\tilde{v})
- p_c \tilde{y}_k (1-\tilde{x}),
\qquad k=1,\dots,r-2
\\
\dot{\tilde{y}}_{r-1} &=
&
p_c \tilde{y}_{r-2}(1-\tilde{x}-\tilde{v})
-
p_c \tilde{y}_{r-1}(1-\tilde{x})
\\
\dot{\tilde{v}} & = &
p_c(1-p)\tilde{x} \tilde{y}
-
p_c q \tilde{v} \tilde{y}
\\
\dot{\tilde{y}}  & = &\sum_{k=0}^{r-1}\dot{\tilde{y}}_k .
\end{eqnarray*}

This system of equations is an extension of the classical model presented in \cite{DK}. It can also be viewed as a variant of the model introduced in \cite{Piqueira2020}. Our goal is not to study this system in detail, but rather to use it to analyse the initial evolution of the process.

 \subsubsection{Early-stage propagation of the fake news}

As in the case $N>1$, we are interested in studying the early stages of rumor propagation in order to determine whether the fake news will die out or spread throughout the population. The point $(x,y_0,\ldots,y_{r-1},v)=(1,0,\ldots,0,0)$ is an equilibrium of the previous system and can therefore be analyzed through a local stability study. However,  this is not the only equilibrium point of the system. Other equilibria also exist, for instance those corresponding to states in which no spreaders remain in the population. In this work, we restrict our attention to the rumor-free equilibrium and do not investigate the stability properties of the other equilibrium points.

The Jacobian matrix computed at this initial equilibrium has a degenerate structure, with several zero eigenvalues. The Jacobian matrix at initial equilibrium  states
\[
J =
\begin{pmatrix}
0 & -p_c  & -p_c  & \cdots & -p_c  & 0 \\
0 & p_c p & p_c  p & \cdots & p_c  p & 0 \\
0 & 0 & 0 & \cdots & 0 & 0 \\
\vdots & \vdots & \vdots & \ddots & \vdots & \vdots \\
0 & 0 & 0 & \cdots & 0 & 0 \\
0 & p_c  (1-p) & p_c (1-p) & \cdots & p_c (1-p) & 0
\end{pmatrix}.
\]
Since the rows corresponding to $\tilde{y}_k$ for $k\ge1$ are zero, the Jacobian has $r-1$ zero eigenvalues.

The presence of multiple zero eigenvalues reflects the degeneracy of the system at the rumor-free equilibrium.
To overcome this issue, we consider the total proportion of spreaders
\[
\tilde{y}(t) = \sum_{k=0}^{r-1} \tilde{y}_k(t).
\]

Using the system equations, we get
\[
\dot{\tilde{y}} = p_c p \tilde{x} \tilde{y} + \alpha (\tilde{y} - \tilde{y}_{r-1})(1 - \tilde{x} - \tilde{v})
- \alpha \tilde{y} (1-\tilde{x}).
\]
Notice that near the equilibrium:
$
\tilde{x} \approx 1, \tilde{v} \approx 0$ and $\tilde{y}_{r-1} \approx 0,$
we get
\[
\dot{\tilde{y}} \approx  p_c p \cdot 1 \cdot \tilde{y} +  p_c (\tilde{y} - 0)\cdot 0 -  p_c \tilde{y} \cdot 0 = p_c p\, \tilde{y}.
\]
Again, we can identify  $R_0 = p_c  p$ as the basic reproduction number in this situation,

In this case, in the early stage of propagation, the spreader population can also be approximated by a Galton--Watson branching process. Now, the offspring distribution is
$
\xi \sim \mathrm{Bernoulli}(p_c p).
$
Indeed, a spreader has a contact during a given time step with probability \(p_c\). Conditional on a contact occurring, the contacted ignorant becomes a spreader with probability \(p\). Hence, a new spreader is generated with probability
$
p_c p.
$
The corresponding basic reproduction number is now
$
R_0 = p_c p.
$

In particular, $R_0=1$ only if $p=1$ and $p_c=1$ (that is, everyone has a contact and there are no checkers). Therefore, in general,
$
R_0 > 1 $ is not attainable, which reflects the absence of an intrinsic epidemic threshold in this 
case. For this reason,  it is not necessary to study the early-stage propagation  with fact-checkers

\section{Simulations}

Although the main objective of this work is to develop an epidemiologically inspired framework for modeling the spread of fake news and to investigate the role of fact-checkers in controlling its propagation, in this section we use simulations to gain further insight into the dynamics of the process.

We will analyze both cases: the situation in which $N>1$, and the case $N=1$ with a contact probability $p_c.$  Recall that $p$ denotes the probability that, upon contact between an ignorant individual and a spreader, the ignorant becomes a spreader. Consequently, $1-p$ represents the probability that the individual becomes a checker.

As is well known in the DK and MT models, two of the most commonly studied parameters are the final proportion of ignorants and the time required for the rumor diffusion process to stop.

Here, we follow a similar approach, but our main objective is to analyze the influence of $p$, which determines whether, after a contact, an individual becomes a spreader or a checker. We present graphs, as a function of $p,$ showing (i) the proportion of the population that ultimately does not believe or does not receive the fake news, and (ii) the duration of the process until this proportion stabilizes. As we have already mentioned, our goal is to analyse the influence of $p$, which controls the number of checkers in the system.

In the first graph, we therefore account for both ignorants and fact-checkers at the moment when rumor propagation ceases. We assume that fact-checkers who transition into stiflers after interacting with a spreader do so because they develop doubts about the validity of the rumor.

The simulations were performed in  R using 100 repetitions, and the results are obtained by averaging the 100 trajectories. Some parameters are kept fixed throughout all simulations: $Y_0 = 4$, meaning that the process starts with 4 spreaders, and $q = 0.25$, which represents the probability that a fact-checker becomes a stifler after interacting with a spreader.

\subsection{The Role of the Initial Proportion of Spreaders}

\begin{figure}[H]
    \centering
    \includegraphics[width=0.8\textwidth]{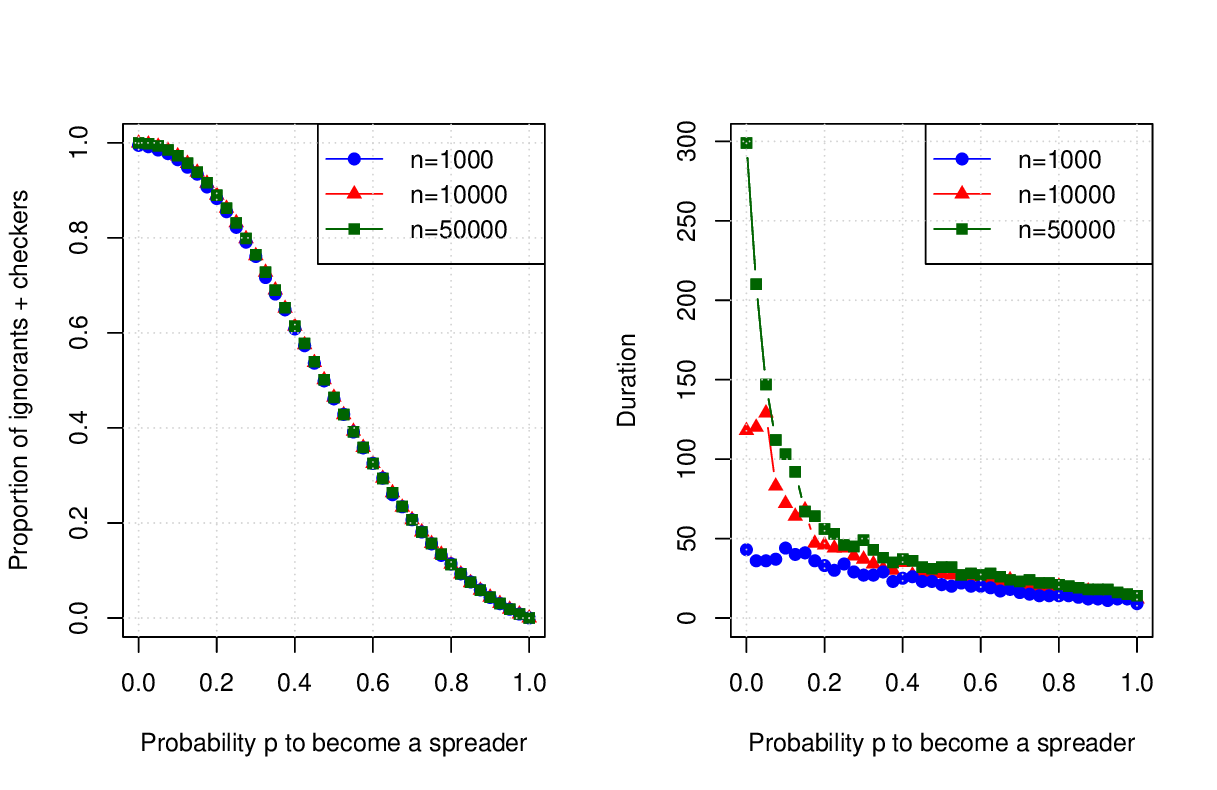}
    \caption{ We consider the case $N=2$ with $r=3$ for different population sizes, always starting with 4 spreaders.
}
    \label{g1}
\end{figure}

\begin{figure}[H]
    \centering
    \includegraphics[width=0.8\textwidth]{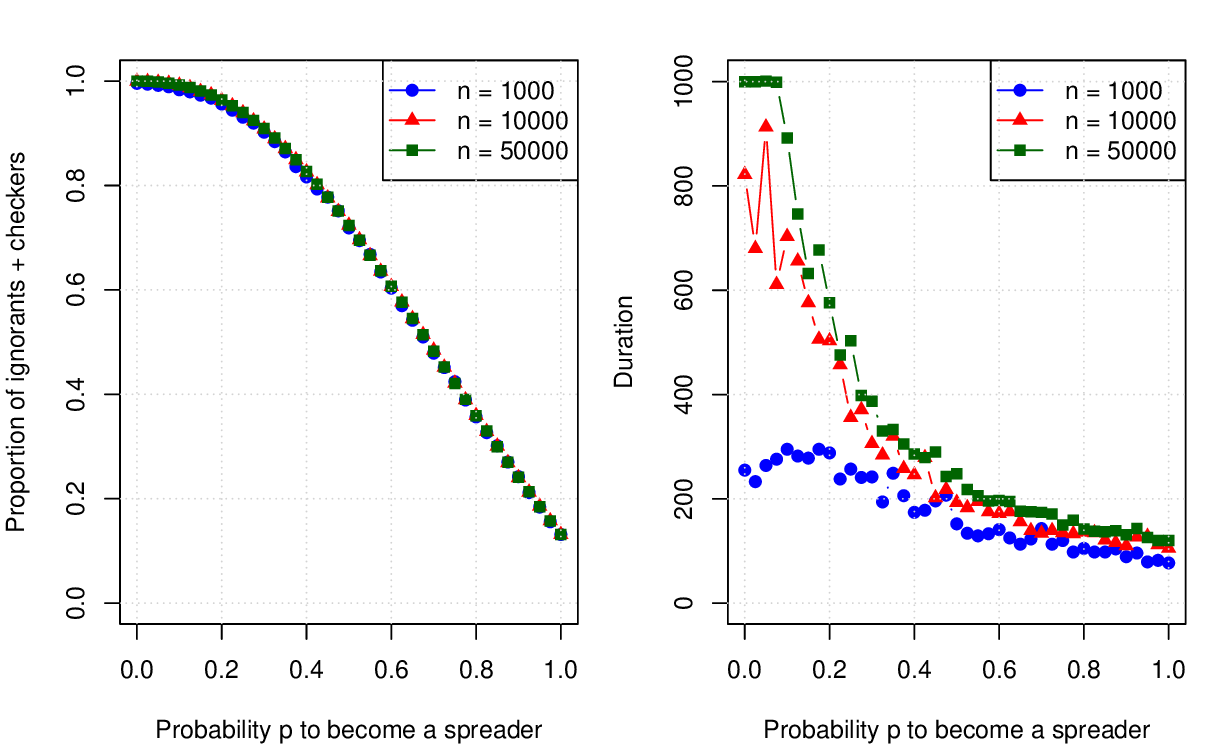}
    \caption{We consider the case $N=1, p_c=0.25$ with $r=2$ for different population sizes.
}
    \label{g2}
\end{figure}

As in the Maki--Thompson model, where the final proportion of ignorants converges to approximately $0.2031$ for large populations independently of the initial proportion of spreaders, a similar phenomenon can be observed in our model, as illustrated in Figures \ref{g1} and \ref{g2}. 

We also observe that the relationship between $p$ and the final proportion of ignorants and checkers is not linear.

\subsection{The Role of the Number of Contacts $N$}

\begin{figure}[H]
    \centering
    \includegraphics[width=0.8\textwidth]{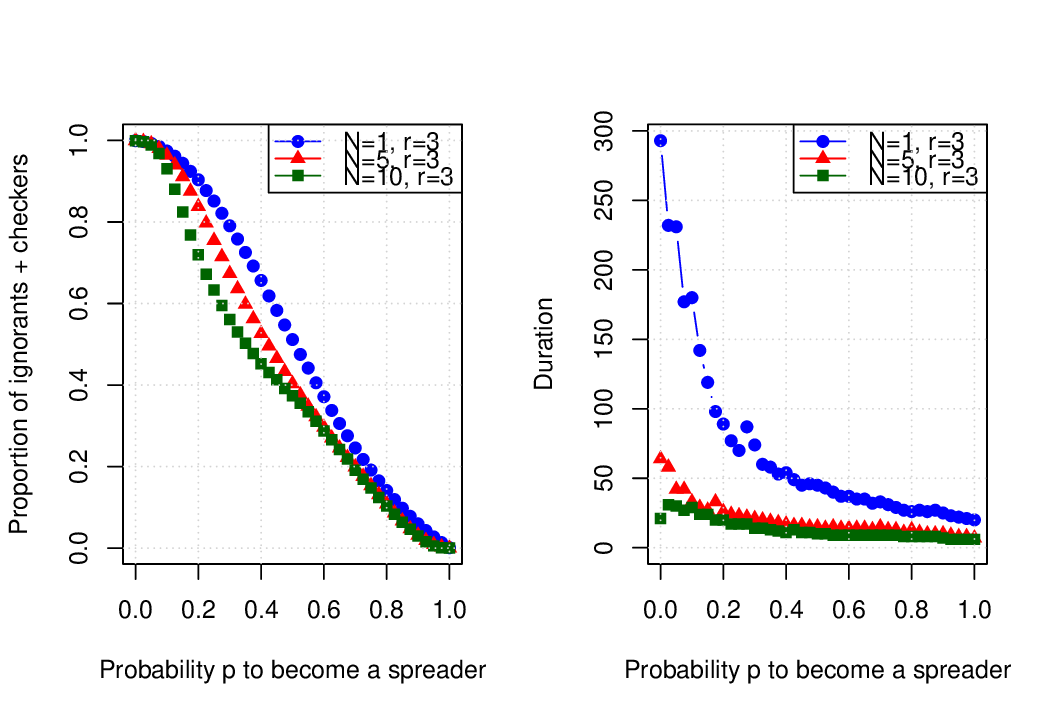}
    \caption{We consider different numbers of contacts $N$, while keeping $r = 3$ fixed.}
    \label{g3}
\end{figure}

\begin{figure}[H]
    \centering
    \includegraphics[width=0.8\textwidth]{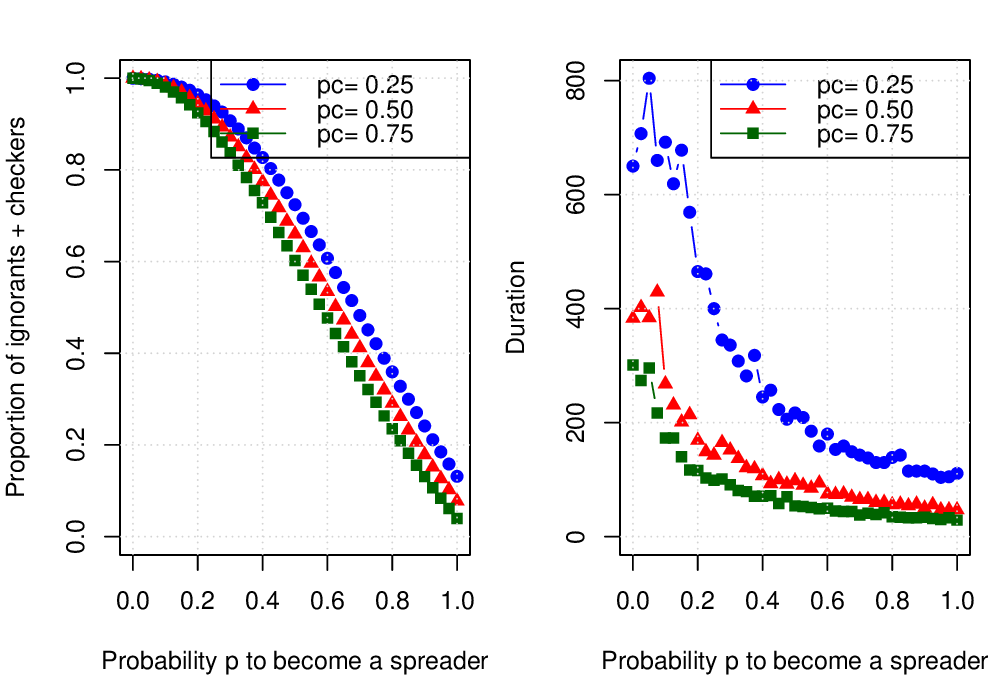}
    \caption{We consider different contact probabilities with $N = 1$, while keeping $r = 2$ fixed.}
    \label{g4}
\end{figure}
In this section, we consider a population of size $n=10000$. We observe, as illustrated in Figures \ref{g3} and \ref{g4},  that the final proportion of ignorants and checkers does not depend significantly on the number of contacts, whereas
 the duration of the process is clearly affected by it.

\subsection{The Role of the Number of Contacts Until Becoming a Non-Spreader $r$}

\begin{figure}[H]
    \centering
    \includegraphics[width=0.8\textwidth]{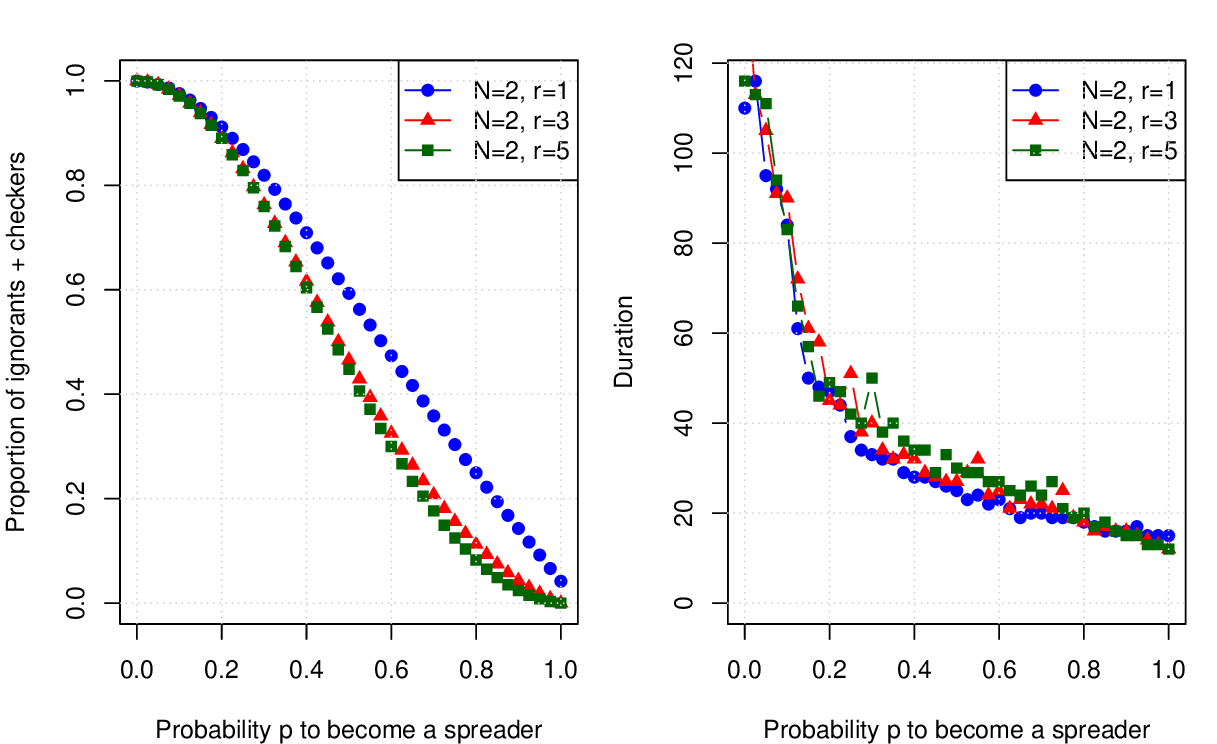}
    \caption{We consider different contact probabilities with $N = 1$, while keeping $r = 2$ fixed.}
    \label{g5}
\end{figure}

\begin{figure}[H]
    \centering
    \includegraphics[width=0.8\textwidth]{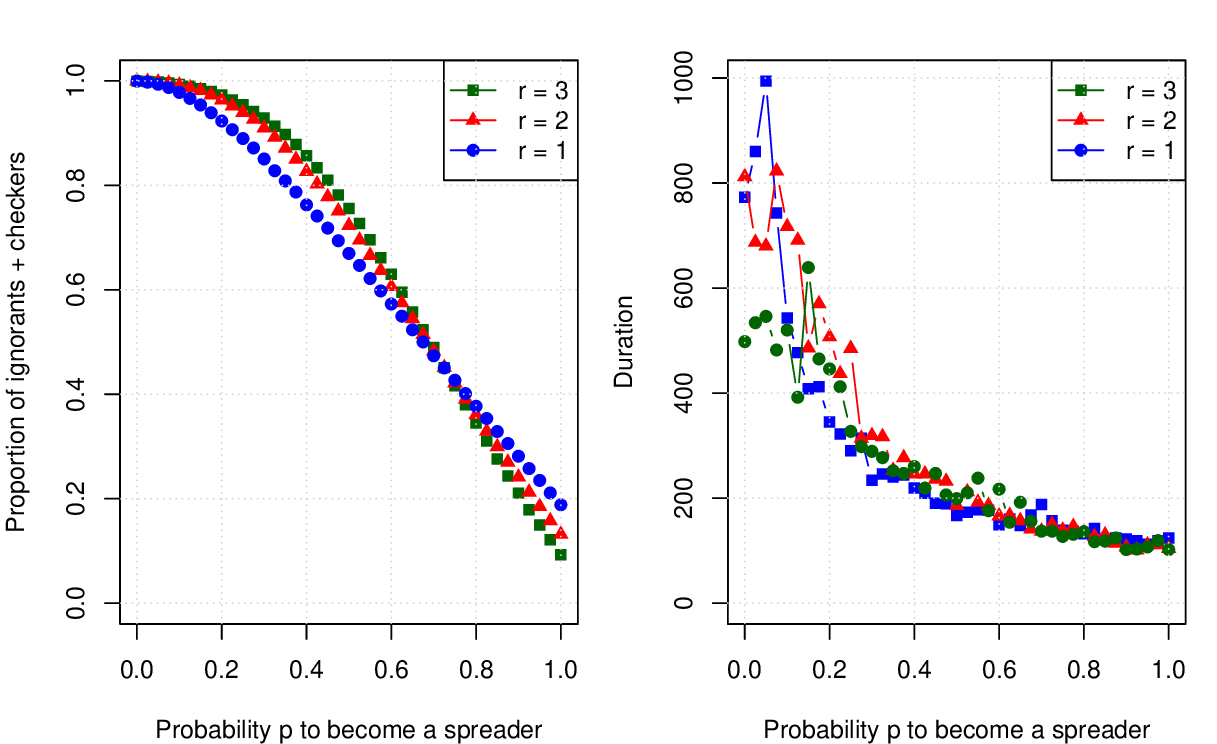}
    \caption{ We consider different values of $r$ with $N = 1$ and a fixed contact probability $p_c = 0.25$.}
    \label{g6}
\end{figure}
In this section, we also consider a population of size $n=10000$. We observe, in Figures \ref{g5} and \ref{g6}, that the final number of ignorants and checkers does not depend significantly on the parameter $r$.

\section{Conclusion}

Our results show that the most effective mechanism for limiting the spread of fake news is the presence of a  large proportion of fact-checkers in the population. When individuals are more likely to verify information rather than propagate it, the final proportion of the population that does not believe or spread the rumor (ignorants plus checkers) increases significantly.
Moreover, the relationship between the key parameter $p$ — the probability of becoming a spreader — and the final outcome is clearly non-linear. 
Importantly, these qualitative behaviors remain robust across all scenarios analyzed, including different population sizes and interaction mechanisms. This indicates that the dominant role of fact-checkers and the non-linear nature of the system are intrinsic properties of the model.

From a practical perspective, the results suggest that increasing the proportion of individuals who verify information before sharing it may be one of the most effective strategies for limiting the spread of fake news. This finding is consistent with current efforts by  independent fact-checking organizations, which aim to encourage critical evaluation of online content.

Despite its usefulness, the model relies on several simplifying assumptions. For instance, the population is assumed to be homogeneous, all individuals share the same behavioral parameters, and fact-checkers are assumed to identify misinformation correctly. Moreover, the model does not account for the structure of real social networks or external information sources. While these assumptions allow for a tractable mathematical analysis, future work could incorporate more realistic interaction networks and heterogeneous individual behaviors.

\end{document}